\documentclass[12pt]{article}
\usepackage{float}  
\usepackage{subcaption}
\usepackage{amsmath, amssymb, graphicx, natbib, geometry}
\title{A Mean-Reverting Model of Exchange Rate Risk Premium Using Ornstein-Uhlenbeck Dynamics}
\author{SeungJae Hwang}
\date{\today}

\begin{document}
\maketitle

\begin{abstract}
This paper proposes a mean-reverting model of the exchange rate risk premium, motivated by persistent empirical deviations from the uncovered interest parity (UIP) condition. We define a realized risk premium as the difference between the observed exchange rate change and the interest rate differential, and demonstrate that this premium exhibits strong mean-reversion across multiple forecast horizons. Modeling the premium as an Ornstein-Uhlenbeck (OU) process embedded within a stochastic differential equation for the exchange rate, we derive analytical expressions for the predictive distribution of future exchange rates.

Empirical analysis using USD/KRW data from 2010 to 2025 shows that the model performs well across both short and long horizons. Forecast coverage is particularly strong at the 2-week and 1-month intervals, and remains robust at 6 months and 1 year, though the 1-year tails exhibit signs of overconservatism. The 3-month horizon consistently underperforms, indicating a transitional regime not fully captured by the single-process specification.

These findings suggest that exchange rate deviations from UIP may stem from a structured, mean-reverting risk premium rather than from random shocks. The proposed model provides a tractable and interpretable framework that links short-term oscillations with long-run convergence, while also identifying key areas\textemdash such as tail calibration and regime dynamics\textemdash for future refinement.
\end{abstract}

\section{Introduction}

The uncovered interest parity (UIP) condition is a foundational principle in international finance, stating that the expected change in the exchange rate should equal the interest rate differential between two countries:

\[
\mathbb{E}_t\left[ \log \left( \frac{S_{t+h}}{S_t} \right) \right] = i_{US,t} - i_{KR,t}.
\]

Although theoretically appealing, UIP has been shown to fail consistently in empirical studies \citep{fama1984forward}. These deviations are commonly attributed to a time-varying risk premium, which may reflect investor sentiment, macroeconomic uncertainty, or global liquidity cycles.

This paper begins with the hypothesis that deviations from UIP may reflect random noise but investigates whether they instead follow a structured, mean-reverting process. To this end, we define a realized risk premium:

\[
K_t := \log \left( \frac{S_{t+h}}{S_t} \right) - (i_{US,t} - i_{KR,t}),
\]

and observe that $K_t$ exhibits mean-reversion across various horizons. This empirical regularity motivates modeling the premium as an Ornstein-Uhlenbeck (OU) process embedded within a stochastic differential equation (SDE) for the exchange rate.

\begin{figure}[H]
\centering
\begin{subfigure}{0.45\textwidth}
\includegraphics[width=\linewidth]{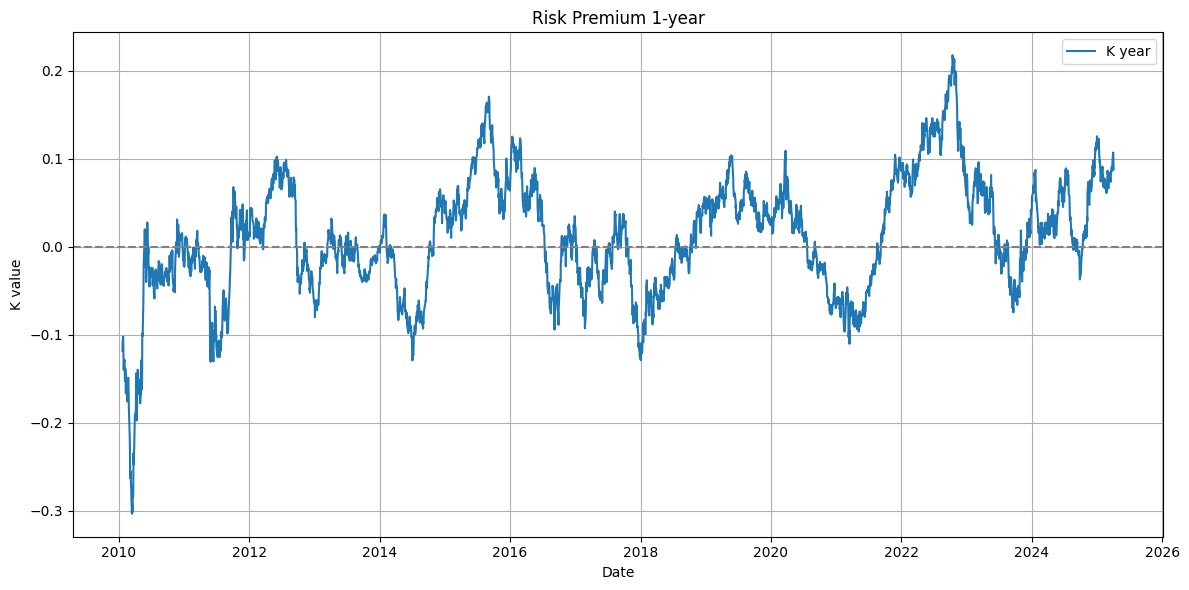}
\caption{1-Year Horizon}
\end{subfigure}
\begin{subfigure}{0.45\textwidth}
\includegraphics[width=\linewidth]{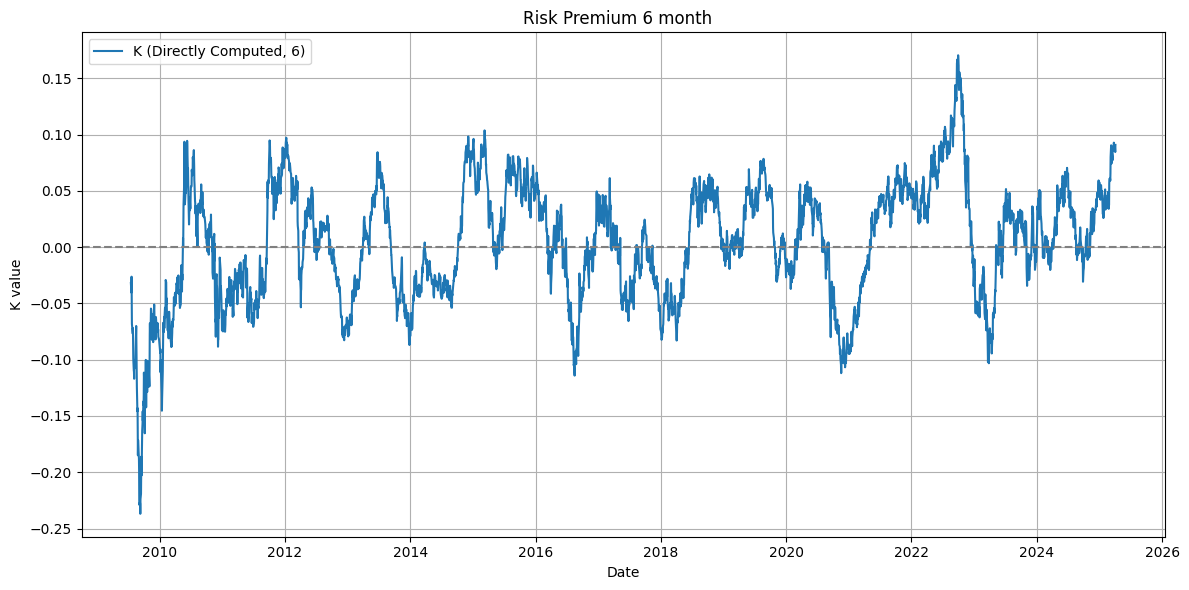}
\caption{6-Month Horizon}
\end{subfigure}
\vspace{0.5em}
\begin{subfigure}{0.45\textwidth}
\includegraphics[width=\linewidth]{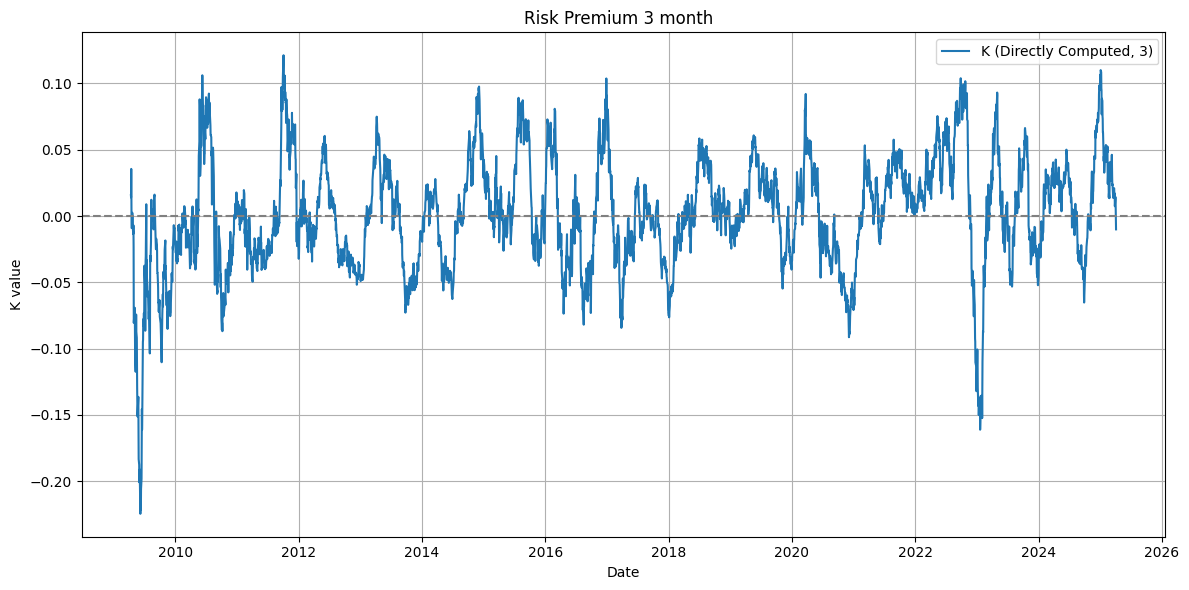}
\caption{3-Month Horizon}
\end{subfigure}
\begin{subfigure}{0.45\textwidth}
\includegraphics[width=\linewidth]{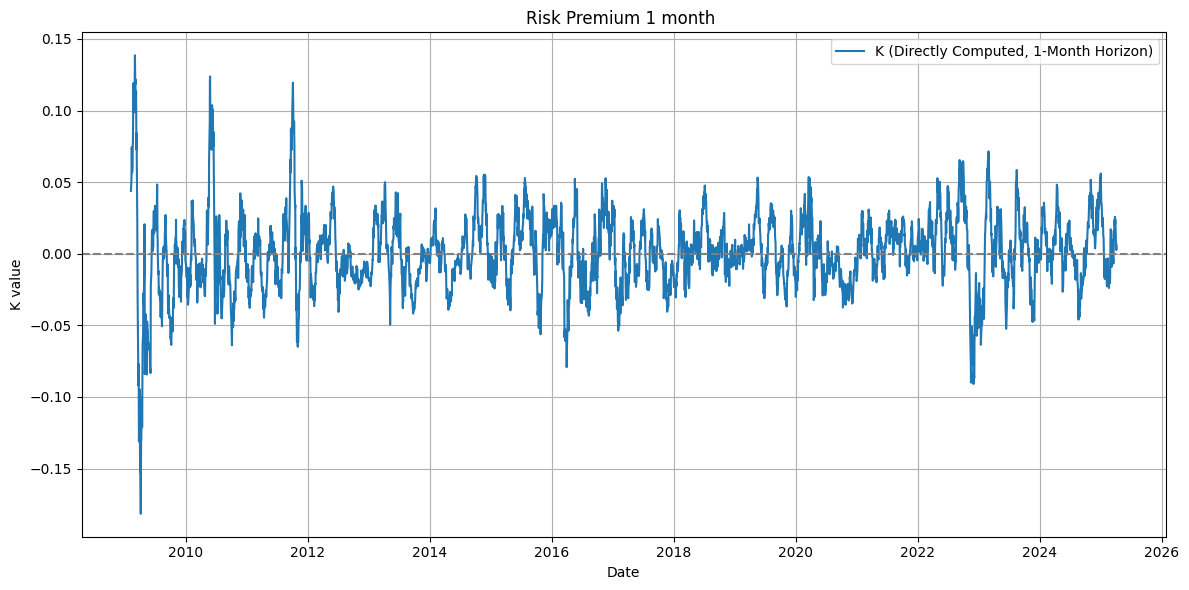}
\caption{1-Month Horizon}
\end{subfigure}
\vspace{0.5em}
\begin{subfigure}{0.45\textwidth}
\includegraphics[width=\linewidth]{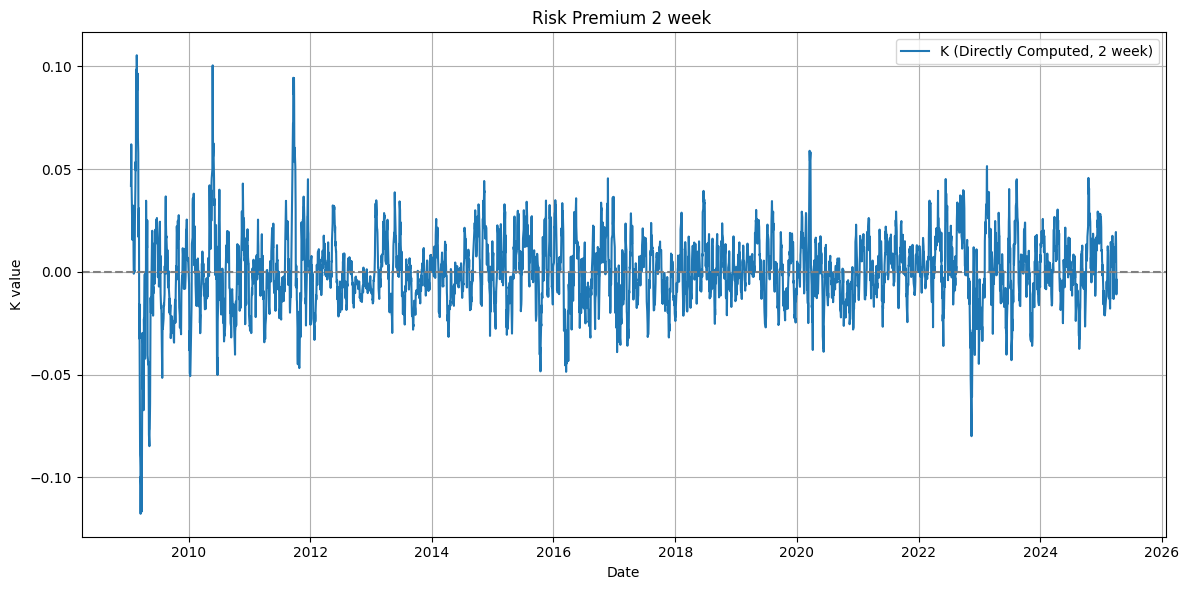}
\caption{2-Week Horizon}
\end{subfigure}
\caption{Estimated risk premium series $K_t$ for different forecasting horizons. All exhibit mean-reverting behavior around zero, with shorter horizons displaying lower volatility.}
\label{fig:kt_horizons}
\end{figure}

We model this behavior using the following OU process:

\begin{equation}
dK_t = \theta(\mu - K_t)dt + \sigma_K dZ_t,
\end{equation}

a continuous-time mean-reverting stochastic process widely used in financial modeling \citep{oksendal2003sde}. Incorporating $K_t$ into the exchange rate dynamics leads to the full SDE:

\begin{equation}
\frac{dS_t}{S_t} = (i_{US,t} - i_{KR,t} + K_t)dt + \sigma_S dW_t,
\end{equation}

from which we derive analytical approximations for the future distribution of exchange rates. This enables closed-form forecasting over multiple horizons ranging from 2 weeks to 1 year.

A central innovation of this study lies in our evaluation framework. Instead of point forecasting, we assess predictive performance using coverage-based backtesting---measuring how often realized exchange rates fall within the model’s confidence intervals. Applied to USD/KRW data from 2010 to 2025, the model shows strong empirical performance across most horizons.

At short-term intervals (2 weeks and 1 month), the model captures exchange rate movements tightly within forecast bands, reflecting the strong impact of short-term mean reversion. It performs surprisingly well even at the 6-month and 1-year horizons, though high coverage at upper confidence levels in the 1-year case (e.g., 99.98\% at 95\%) may indicate overestimated tail uncertainty due to the asymptotic flattening of the OU variance. This suggests that while long-term convergence is captured, the predictive distribution’s tails could be refined.

The 3-month horizon consistently underperforms, with coverage falling below nominal levels across all intervals. We interpret this as a structural transition zone, where short-term reversion has weakened but long-term stability has not yet emerged---possibly due to latent macro shocks or regime shifts. This highlights a potential limitation of the single-regime OU model in capturing transitional dynamics.

In sum, while this research initially assumed that the risk premium would vanish in the long run, our findings reveal a persistent and structured mean-reverting behavior. Modeling this process explicitly yields a probabilistic framework that performs well across most horizons and offers a new lens through which to understand time-dependent deviations from UIP. It also highlights specific weaknesses---notably in the 1-year tails and 3-month transition---that point to concrete directions for future model refinement.

\section{Model Framework}
With the risk premium, the exchange rate dynamic becomes:
\begin{equation}
\frac{dS_t}{S_t} = (i_{US,t} - i_{KR,t} + K_t)dt + \sigma_S dW_t
\end{equation}
We model $K_t$ using the Ornstein-Uhlenbeck (OU) process:
\begin{equation}
dK_t = \theta(\mu - K_t)dt + \sigma_K dZ_t
\end{equation}
Here, $W_t$ and $Z_t$ are independent standard Brownian motions.

\section{Derivation of Exchange Rate Dynamics}
The system \{ $\log S_t$, $K_t$ \} forms a coupled system of stochastic differential equations (SDEs). Since $K_t$ follows an Ornstein-Uhlenbeck (OU) process, it admits an exact analytical solution:
\begin{equation}
K_t = K_0 e^{-\theta t} + \mu (1 - e^{-\theta t}) + \sigma_K \int_0^t e^{-\theta(t - s)} dZ_s
\end{equation}
Substituting $K_s$ into the expression for $\log S_t$ yields:
\begin{equation}
\log S_t = \log S_0 + \int_0^t (r_{int} + K_s) ds + \sigma_S W_t
\end{equation}
Expanding the integral using the analytical form of $K_s$, the deterministic part becomes:
\begin{align}
\int_0^t K_s ds &= \int_0^t \left[ K_0 e^{-\theta s} + \mu (1 - e^{-\theta s}) \right] ds + \int_0^t \sigma_K \int_0^s e^{-\theta(s - u)} dZ_u ds \\
&= \frac{K_0}{\theta} (1 - e^{-\theta t}) + \mu \left(t - \frac{1 - e^{-\theta t}}{\theta} \right) + \text{noise term}
\end{align}
The third term involving the double stochastic integral is a zero-mean Gaussian process, and although it cannot be simplified in closed form, its distributional properties are well understood. As such, we obtain the following approximation for the log exchange rate:
\begin{align}
\log S_t \approx \log S_0 &+ r_{int} t + \frac{K_0}{\theta}(1 - e^{-\theta t}) + \mu\left(t - \frac{1 - e^{-\theta t}}{\theta}\right) \\
&+ \sigma_S W_t + \sigma_K \int_0^t e^{-\theta(t - s)} dZ_s
\end{align}
This expression illustrates that $\log S_t$ is approximately normally distributed. The approximation is valid because the deterministic terms are exact, and the stochastic component, while not expressible in closed form, has a known mean and variance.

\subsection*{Approximation of $S_t$ Distribution via PDF}
Since $\log S_t$ is approximately normal, we can approximate the probability density function (PDF) of $S_t$ using the log-normal distribution. Let $\mu_t$ and $\sigma_t^2$ be the mean and variance of $\log S_t$ computed from the above expression. Then the PDF of $S_t$ is given by:
\begin{equation}
    f_{S_t}(s) = \frac{1}{s \sqrt{2\pi \sigma_t^2}} \exp\left( - \frac{(\log s - \mu_t)^2}{2 \sigma_t^2} \right), \quad s > 0
\end{equation}
To approximate the variance $\sigma_t^2$ of $\log S_t$, we sum the variances from the independent components:
\begin{equation}
\text{Var}[\log S_t] \approx \sigma_S^2 t + \frac{\sigma_K^2}{2\theta} (1 - e^{-2\theta t})
\end{equation}
This expression arises from the variance of the Brownian motion term $\sigma_S W_t$ and the variance of the OU-driven integral. It captures the increasing uncertainty in the short run due to the OU process, which gradually levels off due to its mean-reverting nature.

\subsection*{Derivation of OU Integral Variance}
To justify the variance approximation, consider the OU integral component:
\begin{equation}
X_t := \int_0^t e^{-\theta(t - s)} dZ_s
\end{equation}
Then the variance of $X_t$ is given by:
\begin{align}
\text{Var}[X_t] &= \int_0^t e^{-2\theta(t - s)} ds = e^{-2\theta t} \int_0^t e^{2\theta s} ds \\
&= e^{-2\theta t} \cdot \left( \frac{1}{2\theta} e^{2\theta t} - \frac{1}{2\theta} \right) = \frac{1 - e^{-2\theta t}}{2\theta}
\end{align}
Hence,
\begin{equation}
\text{Var}[\log S_t] = \sigma_S^2 t + \sigma_K^2 \cdot \frac{1 - e^{-2\theta t}}{2\theta}
\end{equation}
The variance of the OU integral is derived using standard results from stochastic calculus (see Øksendal 2003).
\section{Data and Estimation}
\begin{itemize}
  \item \textbf{Data period}: January 2010 to April 2025
  \item \textbf{Exchange rate}: USD/KRW daily spot rates (close prices)
  \item \textbf{Interest rates}: 10-year government bond yields for the US and South Korea
  \item \textbf{Risk premium construction}: $K_t = \log\left( \frac{S_{t+n}}{S_t} \right) - (i_{US,t} - i_{KR,t})$
  \item \textbf{Motivation}: The constructed $K_t$ series displays mean-reverting behavior, motivating the use of the Ornstein-Uhlenbeck (OU) process
  \item \textbf{Estimation method}: OU parameters $(\theta, \mu, \sigma_K)$ estimated via maximum likelihood estimation (MLE)
\end{itemize}

\section{Backtesting Methodology and Results}

To evaluate the forecasting performance of our stochastic model, we conduct a backtesting exercise across various horizons: 2 weeks (10 business days), 1 month (21 days), 3 months (63 days), 6 months (126 days), and 1 year (252 days). We use daily data on the USD/KRW exchange rate and interest rates from January 2010 to April 2025. The dataset is divided into a training set (80\%) and a validation set (20\%), with the latter used for out-of-sample evaluation.

At each forecast origin $t$, we compute the expected log exchange rate at horizon $t + h$ using the derived model:
\begin{equation}
\mathbb{E}_t[\log S_{t+h}] = \log S_t + r_{int} + \frac{K_t}{\theta}(1 - e^{-\theta h}) + \mu \left(h - \frac{1 - e^{-\theta h}}{\theta}\right)
\end{equation}

The variance of the forecast distribution is given by:
\begin{equation}
\text{Var}[\log S_{t+h}] = \sigma_S^2 h + \frac{\sigma_K^2}{2\theta}(1 - e^{-2\theta h})
\end{equation}

Given the approximate log-normality of $S_{t+h}$, confidence intervals for future spot rates are constructed as:
\begin{equation}
[\exp(\mu_t - z \cdot \sigma_t),\ \exp(\mu_t + z \cdot \sigma_t)]
\end{equation}
where $\mu_t$ and $\sigma_t^2$ are the mean and variance above, and $z$ is the corresponding quantile (e.g., $z = 1.96$ for a 95\% level).

We compute the coverage rate, defined as the fraction of actual future exchange rates that fall within the model’s predicted confidence interval. The results are summarized below:

\subsection*{Coverage Rates Across Horizons (Validation Set)}

We report empirical coverage rates for confidence intervals over 2-week, 1-month, 3-month, 6-month, and 1-year horizons:

\begin{table}[H]
\centering
\begin{tabular}{|c|c|c|c|c|c|}
\hline
Confidence Level & 2-Week & 1-Month & 3-Month & 6-Month & 1-Year \\
\hline
50\% & 52.95\% & 56.90\% & 36.90\% & 54.87\% & 48.88\% \\
60\% & 63.49\% & 66.19\% & 47.61\% & 66.73\% & 55.27\% \\
70\% & 73.40\% & 74.32\% & 58.32\% & 74.69\% & 63.90\% \\
80\% & 80.68\% & 82.84\% & 66.43\% & 82.65\% & 75.72\% \\
90\% & 91.22\% & 92.13\%  & 75.11\% & 90.44\% & 92.97\% \\
95\% & 95.73\% & 96.77\% & 81.91\% & 95.58\% & 99.98\% \\
99\% & 98.24\% & 98.71\% & 89.15\% & 97.88\% & 100\% \\
\hline
\end{tabular}
\caption{Empirical coverage rates by forecast horizon and nominal confidence level.}
\end{table}

The model exhibits strong coverage performance, particularly at short-term horizons (2 weeks and 1 month), where empirical coverage closely aligns with theoretical confidence levels. This finding supports the interpretation that short-term deviations from uncovered interest parity (UIP) are not purely random, but instead reflect systematic and predictable risk premia.
\section{Empirical Results and Discussion}

We evaluate the predictive accuracy of the proposed model across five forecast horizons---2 weeks, 1 month, 3 months, 6 months, and 1 year---by comparing the empirical coverage rates of predicted confidence intervals against realized exchange rate movements. This probabilistic approach allows us to assess how well the model accounts for uncertainty and directional movements across time scales.

At short-term horizons (2 weeks and 1 month), the model performs strongly. The risk premium quickly reverts toward its mean, resulting in narrow confidence intervals that capture realized exchange rate movements with high accuracy. Coverage rates slightly exceed nominal levels, indicating a conservative but reliable forecast. These findings support the interpretation that short-term deviations from uncovered interest parity (UIP) are not merely noise, but reflect structured, mean-reverting risk premia.

The model also performs well at the 6-month horizon, with coverage rates closely aligned with target confidence levels. This suggests that the mean-reverting structure remains informative even at medium-to-long horizons, capturing the gradual convergence dynamics of the exchange rate.

However, the 3-month horizon presents a consistent anomaly. Across all confidence levels, coverage rates are significantly lower than expected. This suggests that the 3-month forecast window may represent a transitional regime---a zone where short-term reversion has weakened, but long-term equilibrium forces have not yet taken hold. Market noise, policy uncertainty, or macroeconomic shocks may exert disproportionate influence in this interval. From a modeling standpoint, this highlights the limitations of a single-regime Ornstein-Uhlenbeck (OU) process and points to the potential need for regime-switching or non-linear dynamics in the mid-horizon context.

The 1-year horizon produces a distinct pattern. While lower confidence levels (50\% to 80\%) yield coverage rates that are reasonably close to nominal, higher confidence intervals (95\% and 99\%) display near-perfect coverage, reaching 99.98\% and 100\%, respectively. While this might seem desirable, it more likely reflects an overestimation of long-term uncertainty due to the OU process’s asymptotically stabilizing variance. These wide predictive intervals may trivially enclose realized outcomes, particularly in the tails. Thus, the model’s long-horizon density forecast appears overly conservative in its extremes, despite reasonable calibration in the central region. This discrepancy reveals an important avenue for improvement: refining the model’s predictive density structure at long horizons through tail-sensitive distributional forms or variance adjustment mechanisms that better reflect long-run bounded risk.

In sum, the empirical findings demonstrate that the proposed mean-reverting risk premium model captures key aspects of exchange rate dynamics, especially in the short and long term. Nonetheless, its underperformance at the 3-month horizon and overconservatism in the far-tail regions at the 1-year horizon suggest meaningful opportunities for further refinement. Incorporating dynamic volatility structures, regime shifts, or macroeconomic covariates may enhance the model’s ability to characterize transitional behaviors and long-term tail risks more effectively.

\section{Conclusion}

This paper set out to investigate whether deviations from uncovered interest parity (UIP) reflect a temporary risk premium that fades over time. Starting from the hypothesis that the premium should vanish in the long run---allowing exchange rates to eventually revert to parity---we empirically uncovered strong and persistent mean-reverting behavior in the estimated premium across multiple horizons.

Motivated by this empirical regularity, we modeled the risk premium as an Ornstein-Uhlenbeck (OU) process and embedded it within a continuous-time stochastic model of the exchange rate. The resulting framework produced analytical forecasts with time-varying variance, enabling quantitative evaluation of predictive performance.

Backtesting results on USD/KRW data from 2010 to 2025 demonstrated strong performance at both short-term and long-term horizons, despite our initial expectation that predictability would emerge only in the long run. Notably, the model underperformed at the 3-month horizon---a result we interpret as stemming from transitional volatility, where short-term reversion has not fully taken effect and long-run convergence is not yet dominant.

These findings suggest that exchange rate risk premia are not simply residual noise or market inefficiencies, but structured components of exchange rate dynamics that can be explicitly modeled and predicted. Our work contributes a tractable framework that connects short-run fluctuations with long-run equilibrium through a dynamic, mean-reverting premium.

Future research may extend this model by incorporating regime-switching dynamics, macroeconomic variables, or time-varying parameters to better capture transitional phases like the 3-month horizon. Nevertheless, our results affirm the value of treating risk premia as structured, predictable forces in currency markets---and highlight the potential of combining stochastic modeling with empirical validation to improve understanding and forecasting of exchange rate behavior.

\bibliographystyle{apalike}  % 혹은 plainnat, ieee 등 원하는 스타일
\bibliography{references}  

\begin{thebibliography}{}

\bibitem[Fama, 1984]{fama1984forward}
Fama, E.~F. (1984).
\newblock Forward and spot exchange rates.
\newblock {\em Journal of Monetary Economics}, 14(3):319--338.

\bibitem[Øksendal, 2003]{oksendal2003sde}
Øksendal, B. (2003).
\newblock {\em Stochastic Differential Equations: An Introduction with Applications}.
\newblock Springer, 6th edition.

\end{thebibliography}

\end{document}